# Open Gas-Cell Transmission Electron Microscopy at 50 pm Resolution.

Idan Biran[1*], Frederik Dam[1], Sophie Kargo Kaptain[1], Ruben Bueno Villoro[1], Maarten Wirix[2], Christian Kisielowski[3], Peter C. K. Vesborg[1,4], Jakob Kibsgaard[1,4], Thomas Bligaard[1,5], Christian D. Damsgaard[1,4,6], Joerg R. Jinschek[1,6], Stig Helveg[1*]

1. Center for Visualizing Catalytic Processes (VISION), Department of Physics, Technical University of Denmark, DK-2800 Kgs. Lyngby, Denmark.
2. Materials and Structural Analysis Division, Thermo Fisher Scientific, Eindhoven, Netherlands.
3. Electron Scattering Solutions, 337 Moraga Ave, Piedmont CA 94611, USA.
4. SURFCAT, Department of Physics, Technical University of Denmark, DK-2800 Kgs. Lyngby, Denmark.
5. Department of Energy Conversion and Storage, Technical University of Denmark, DK-2800 Lyngby, Denmark
6. National Centre for Nano Fabrication and Characterization (DTU Nanolab), Technical University of Denmark, DK-2800 Kgs. Lyngby, Denmark.

## Abstract:

Transmission electron microscopy (TEM) has reached ~ 50 picometer resolution in a high vacuum, enabling single-atom sensitive imaging of nanomaterials. Extending this capability to gaseous environments would allow for similar visualizations of nanomaterial dynamics under chemically reactive conditions. Here, we examine a new TEM system that maintains 50 pm resolution at pressures up to 1 mbar, demonstrated using nanocrystalline Au immersed in $N_2$. The system features an open gas-cell with a four-stage differential pumping system, a 5th order aberration corrector for broad-beam TEM, a monochromatized electron beam, an ultra-stable microscope platform, Nelsonian low electron dose-rate illumination, and direct electron detection. Young fringe experiments and exit wave phase imaging confirm the atomic resolution and indicate location-dependent vibrational blur at surface terminations. Thus, this platform advances *in situ* and *operando* TEM studies of gas-surface interactions in diverse fields, including catalysis, corrosion, and crystal growth.

## Introduction:

Over the past two decades, transmission electron microscopy (TEM) advanced rapidly to visualize matter at the atomic scale. In particular, improvements for the correction of electron optical aberrations, suppression of incoherences, detection efficiency and mechanical stabilization of microscopes' construction have been pivotal for improving imaging contrast and resolution. Ultimately, a spatial resolution of 50 pm can now be routinely obtained in both broad beam and focused beam modes.[1,2] This resolution level is commonly considered to be ultimately set by the extent of the atoms' electrostatic potential rather than the electron optics, representing a change-over from instrument- to object-limited imaging resolution. Uniquely, it enables for single-atom sensitivity imaging across the Periodic Table of Elements and, in turn, the determination of the three-dimensional atomic structure of nanostructured materials.[3,4]

However, caution is needed in interpreting atomically resolved electron microscopy images of nanomaterials as their surfaces tend to adapt dynamically to the surrounding gaseous environment, reflecting that adsorption and cohesive energies are often comparable in magnitude. Hence, it remains elusive whether the atomic-resolution TEM observations of a nanomaterial obtained under high vacuum conditions are indeed representative of its functional state. This has triggered the advancement of open and closed gas-cells for confining reactive gas environments to the vicinity of the nanomaterial sample inside the vacuum of electron microscopes.[5] Both gas-cells have shown imaging capabilities reaching ~ 100 pm of resolution at pressures of a few mbars to bars[6–9]. Hence, a crucial question remains: Can an inherent object-limited resolution of ca. 50 pm be maintained for nanomaterials during exposure to reactive gas conditions to benefit from single-atom imaging sensitivity in the exploration of gas-surface interactions on nanomaterials' surfaces?

Here we address the question of image resolution under gas environments in TEM mode. A new transmission electron microscope, VISION PRIME, combines for the first time electron optics designed for 50-pm image resolution in broad-beam mode with an open gas-cell capable of confining a gas atmosphere in the mbar range in close vicinity of the specimen. Specifically, the attainable information limit of the microscope is demonstrated by Young fringe (YF) experiments and by exit wave image reconstructions of a nanocrystalline Au sample in high vacuum and under exposure to $N_2$ at a pressure of 1 mbar. While the YF experiments provide an area-integrated assessment of the resolution, the exit wave phase images measure resolution locally at the individual atomic columns uncovering spatial heterogeneity across a nanostructured material.

# Experimental and Methods

1. *VISION PRIME electron microscope*

The VISION PRIME electron microscope is based on Thermo Fisher Scientific's SPECTRA Ultra platform, combining electron optics and detection for reaching 50-pm image resolution in high-resolution transmission electron microscopy (HRTEM) in combination with an open gas-cell.

The microscope is equipped with a high-brightness Schottky field emission gun (XFEG) and a Wien-filter monochromator for generating an electron beam with an energy full-width-half-maximum of $\Delta E = 0.1$ eV at a primary electron energy of $E_0 = 300$ keV in broad-beam mode. The condenser aperture (C1) in the energy-selecting plane improves energy resolution to extend the HRTEM information limit as well as enabling Nelsonian illumination to match the field-of-illumination and field-of-view.[10] Specifically, Nelsonian illumination is herein the combination of a monochromatized beam (with a monochromator excitation value of 1.2) and an inserted circular C1 aperture corresponding to an field-of-illumination area with a radius of 50 nm. The field-of-view, captured by the imaging camera, is set by the projection system to be 39 nm x 39 nm. Moreover, shifting the monochromatic beam across the inserted C1 aperture enables even illumination over the field-of-view as well as fine-tuning the electron dose-rate traversing the microscope. The Nelsonian illumination mode was used to obtain all images reported herein.

The microscope includes a constant power objective lens (TWIN PRIME) for fast switching of the primary electron energy. It has an inherent spherical aberration coefficient $C_s = 1.2$ mm and a chromatic aberration coefficient $C_c = 1.6$ mm (at 300 keV). For this setup, the fractional root-mean-square (rms) ripple in the objective lens current is $\Delta(I)/I \leq 0.2$ ppm rms, and high tension ($V = E_0/e$ with e, the elementary charge) is $\Delta(V)/V \leq 0.2$ ppm rms. Thus, the theoretical upper limit of the focal spread at a primary electron energy of 300 keV is

$$\delta = C_C \left[ 4\left(\frac{\Delta I}{I}\right)^2 + \left(\frac{\Delta E}{E_0}\right)^2 + \left(\frac{\Delta V}{V}\right)^2 \right]^{\frac{1}{2}} \leq 9.1 \text{ Å}$$

The microscope includes a new hexapole-based electron optical aberration-corrector, CETCOR PRIME, which compensates for the optical aberrations of the lower polepiece of the objective lens up to the 5th order from 30-300 kV. It is an improvement from the previous generation as it corrects the six-fold astigmatism such that the transmitted phase space is substantially enhanced for an improved information transfer from ca. 60 pm to ca. 50 pm. Further, it minimizes off-axial four-fold astigmatism and off-axial star aberration. Aberrations were corrected by the Zemlin tableau method,[11,12] in which HRTEM images are acquired using a bottom-mounted metal-oxide semiconductor (CETA-S) camera while the microscope projection system operated at an image pixel size of 33 pm. From the acquisitions taken using the image corrector software, the software calculated fast Fourier transforms (FFTs) and fits the low-order aberration: defocus $C_1$ and two-fold astigmatism $A_1$, while other aberration coefficients are calculated by using an extended version of the Zemlin tableau method. The aberration correction is followed by setting up the first crossover in the contrast transfer function (CTF) at the highest frequency possible. A strong phase contrast is obtained by optimizing the $C_1$, $C_3$, and $C_5$ values,[13] which herein corresponds to a slight overfocus of $C_1$ between 4 nm and 6 nm, $C_3$ of ca. -13.5 μm, and $C_5$ of ca. 7.5 mm.

Moreover, the microscope is equipped with a direct electron detection camera (Falcon 4i) for low electron dose-rate and dose imaging, and a multi-signal data acquisition and analysis platform (Velox, Thermo Fisher Scientific) is used for acquisition and online data processing.

The microscope is also equipped with a post-column energy filter (Gatan Continuum ER 1056, Gatan Inc.) for electron energy loss spectroscopy (EELS). The energy filter was tuned to a zero-loss peak (ZLP) full-width-at-half-maximum of 0.1 eV using TEM image mode with a 2.5 mm entrance aperture to the energy filter and a detector dispersion of 0.15 meV per pixel. With these settings, the ZLP was aligned to 0 eV, and subsequently, electron energy loss spectra were acquired and presented herein without further processing.

Additionally, the microscope's open gas-cell is constructed as a differential pumping system, referred to as a controlled atmosphere electron microscope[14] or environmental transmission electron microscope (ETEM).[15] It includes the first true constant power objective lens modified to establish the first differential pumping stage. Successive differential pumping stages follow conceptually a previous layout,[6] and are added upstream at the objective lens, C3 and C2 condenser apertures, the C1 lens, and the electron gun area, and also downstream at the selected area aperture. Including this four-stage upstream pumping system allows for gas introduction to pressures of up to 20 mbar in the open gas-cell while maintaining the high vacuum conditions at the electron source – a pressure difference of nine orders of magnitude.

The open cell gas handling system is fully software-controlled and consists of eight TMPs, three Scroll vacuum pumps, a series of valves and pressure gauges, a safety interlock system, and a quadrupole mass spectrometer (QMS). It is designed as a separate module mounted on the high-base frame of the SPECTRA Ultra platform to suppress the transfer of mechanical, acoustic, and electromagnetic stray field disturbances to the electron optical performance.

The microscope is mounted on an integrated vibration isolation system (iVIS) that provides active damping of disturbances caused by floor vibrations and is housed inside an enclosure to dampen acoustic and thermal variations from the environment. The entire microscope room is a lab-in-a-lab construction to suppress acoustic disturbances. The inner laboratory rests on a 2-meter thick and ca. 60 m$^2$ wide concrete slab to avoid vibrational transfer from the exterior environment. Also, the inner laboratory is encapsulated in a combined active and passive electromagnetic stray field shield, as designed and installed by Systron EMV.

*2. Information limit estimation using Young fringe images*

Young fringe (YF) images are obtained at a primary electron energy of 300 keV to estimate the information limit of VISION PRIME.[10,12,16] Specifically, YF images were acquired using the Falcon 4i direct electron counting detector (4096 x 4096 pixels) in counting mode, with projector optics matching a pixel size of 8.3 pm/pixel, and by recording and summing up consecutively acquired images of 1-sec exposure each, shifted in x or y directions at a rate of 2 Hz using the image corrector software. The resulting images are fast Fourier transformed, and the information limit is determined by visual inspection as the maximum spatial frequency where the induced periodicity fades out. The YF images were obtained under high vacuum, corresponding to a base pressure of 8 x 10$^{-7}$ mbar with a fully evacuated gas-cell with all pumps and gauges activated (referred to as 0 mbar $N_2$), and $N_2$ leaked into the open gas-cell at 1 mbar. The dose-rate, reported in all measurements, was detected by the direct electron detector, over the sample area of interest.

*3. Focal series acquisition and exit wave reconstruction*

Next, the image resolution is evaluated in real space by an analysis of electron exit wave images. The exit wave images are obtained from a focal series of HRTEM images, acquired using the direct electron detector at similar illumination and environmental conditions employed for the YF experiments. All images in a series were acquired with 4096 x 4096 pixels in counting mode of the direct electron detector, and in immediate succession, spanning 100 images, with a focal step of 1 nm between each image, and a pixel size of 8.3 pm/pixel. Each image in the focal series was acquired with an acquisition time of 0.327 seconds. For focal series acquired with 0 and 1 mbar $N_2$, the electron dose-rate measured over the area of interest was 1200 and 888 e/Å$^2$/sec, respectively.

Following the acquisition, each focal series of images was coarse-aligned by a cross-correlation algorithm using the bandpass filter in DigitalMicrograph (Gatan Inc.) with sub-pixel accuracy.

Subsequently, the aligned image series was used to reconstruct the corresponding complex exit wave using the TEMPAS software,[17] where the complex exit wave is retrieved using an iterative exit wave reconstruction algorithm based on the Gerchberg-Saxton algorithm.[18] The algorithm further corrects any residual image shifts, by phase correlation, and determines the actual defocus values for each of the images in the series, before it reconstructs the electron exit wave. A numerical objective aperture was set at 1/(50 pm), and the exit wave reconstruction is iteratively repeated until the standard deviation of the modulus of a previous and present exit wave is within the limits of one focal step. Residual non-centrosymmetric aberrations are reduced numerically in regions of interest by optimizing column roundness and subsequently maximizing phase values at atomic column positions. From the corrected regions, atomic columns are determined and peak coordinates are logged for further analysis. The dose of a reconstructed exit wave is set by the number of images in the focus series, which was 47 and 63 for the 0 mbar and 1 mbar of $N_2$, respectively, and corresponds to ca. 18700 e/Å$^2$, matching the YF experiments.

 4. *Atomic column width measurements in exit wave phase images*

The exit wave images are used to evaluate the image resolution and its heterogeneity in real space by focusing on crystal domains oriented with a zone-axis along the electron beam. In such regions, the exit wave phase image will have a maximum at the atomic column positions that scales with the atomic content and dependent on the atomic vibrational amplitude.[4,19] The width of the phase peaks can thus be taken as a local measure of resolution in real space.[20–22] The column width is measured following Nord et al.[23] by fitting the exit wave phase image peaks with a generalized two-dimensional (2D) elliptical Gaussian function given by:

$$g(x,y) = A\, exp(-[a(x-x_0)^2 + 2b(x-x_0)^2(y-y_0)^2 + c(y-y_0)^2]) + O$$

where:

$$a = \frac{\cos^2(\theta)}{2\sigma_x^2} + \frac{\sin^2(\theta)}{2\sigma_y^2}, \qquad b = -\frac{\sin(2\theta)}{4\sigma_x^2} + \frac{\sin(2\theta)}{4\sigma_y^2}, \qquad c = \frac{\sin^2(\theta)}{2\sigma_x^2} + \frac{\cos^2(\theta)}{2\sigma_y^2},$$

$O$ is an offset value, $A$ is the amplitude, $x_0$ and $y_0$ are the center positions, $\sigma_x$ and $\sigma_y$ are the standard deviations, and $\theta$ is the rotation angle of the ellipse. For each phase peak, the parameters are fitted within 21 x 21 pixel regions (approximately 175 pm x 175 pm) centered on initial peak coordinates. The fitting area is chosen as a compromise between including a sufficient area for capturing the peak shape and avoiding contributions from neighboring phase peaks. The fitting is performed as follows: Initial values for $A$ and $O$ are the maximum and minimum pixel values present within each region, respectively. Additionally, ($x_0$, $y_0$) are the initial pixel coordinates of the phase peak center. Initial values for $\sigma_x$ and $\sigma_y$ are set to one pixel, and the initial value for $\theta$ is set to be 0. Furthermore, fitting bounds are set to be positive values for A, $\sigma_x$, $\sigma_y$, and O, to the range 0 to $\pi$ for $\theta$ and to the extracted regions of interest for $x_0$ and $y_0$.

The fitting to the atomic columns is represented by the following parameters: (i) An ellipticity value defined as $\varepsilon = \sigma_x/\sigma_y$ for $\sigma_x < \sigma_y$ (and $\varepsilon = \sigma_y/\sigma_x$ for $\sigma_y < \sigma_x$). The ellipticity is $\varepsilon = 1$ for circular symmetric columns. (ii) A rotation angle $\theta$ between the image $x$-axis and the ellipse's major axis for $\sigma_x > \sigma_y$ (and $\theta + \pi/2$ for $\sigma_y > \sigma_x$). (iii) A column width defined as the equivalent diameter $2<\sigma>$ (referred in this work as atomic column width), for which the corresponding circular area ($A = \pi<\sigma>^2$) matches the elliptical area ($A = \pi\sigma_x\sigma_y$). That is, the ultimate resolution is characterized by a perfect single-crystal sample in zone-axis-orientation with $\varepsilon=1$ as $2<\sigma>=2\sigma_x=2\sigma_y$. In practise, deviations are expected for a nanocrystalline sample due to slight sample tilts off the zone-axis orientation, local residual aberrations, contrast modulation by amorphous supporting films, as well as heterogeneous atom vibrations.

# Results and discussion:

The ability to obtain electron micrographs at 50 pm spatial resolution was previously obtained with the specimen being under high vacuum conditions.[1,2] Interfacing such an electron optical system with an open gas-cell adds extra challenges because the gas handling system introduces additional vacuum pumps, valves, pressure gauges, and mass spectrometers that must be sufficiently damped with respect to their vibrational, acoustic, and electromagnetic disturbances.

The successful combination of a 50-pm-resolution TEM with an open gas-cell in the VISION PRIME is demonstrated first by measuring the electron microscope's electron-optical stability with a fully evacuated gas-cell. First, **Figure 1a** shows time-resolved measurements of electron optical aberrations, including the low-order two-fold and three-fold astigmatism ($A_1$ and $A_2$), and coma ($B_2$), as they are less stable than high-order aberrations[24] and key to determine high-order aberrations. The three-minute intervals between each measurement are representative for the acquisition of a time series or focal series of images. **Figure 1a** shows the corresponding aberration value and its confidence level measured repeatedly over a 70-minute period. Specifically, the change in $A_1$ value by less than 2 nm over the full duration of the measurement, with a stable measurement of the angle changing by less than 2°, is a notable improvement, compared to previous studies.[25] **Figure S1** shows the high-order aberrations evaluated from the same experiment as in **Fig. 1a** and reflects a similar high stability that is essential for maintaining the 50 pm resolution. Moreover, additional aberration stability measurements at 30-minutes interval for a 2.5-hour duration are reported in **Fig. S2**. The high stability of the measured aberrations reflects an optical lifetime extended from minutes to hours,[25] indicating a longevity of the optical state that provides well-defined electron-optically data over the typical extent of time-resolved studies of nanomaterials in reactive gas environments.

The VISION PRIME's high temporal stability in the beam direction can be assessed by measurements of the defocus $C_1$ values in a focal series of images. Such measurements convolute the stability of both the electron optics and the sample. **Figure 1b** shows $C_1$ values of a focal series changing linearly from 2.5 nm to 49.5 nm, over a total exposure time of 15 sec, as measured using the Thon ring approach in the TEMPAS software. The difference between any specific image focus value and the best linear fit is defined as the focus error. **Figure. 1b** inset shows the individual focus error values span from -5Å to 5Å, and a Gaussian fit reflects a standard deviation of 2.47Å, for the present measurements. Additionally, the microscope's stability is further supported by electron energy loss spectroscopy (EELS) of the transmitted electron beam. **Figure 1c** shows the position of the ZLP shifts by less than 0.19 eV over 1-hour of consecutive measurements taken in 10-minute intervals.

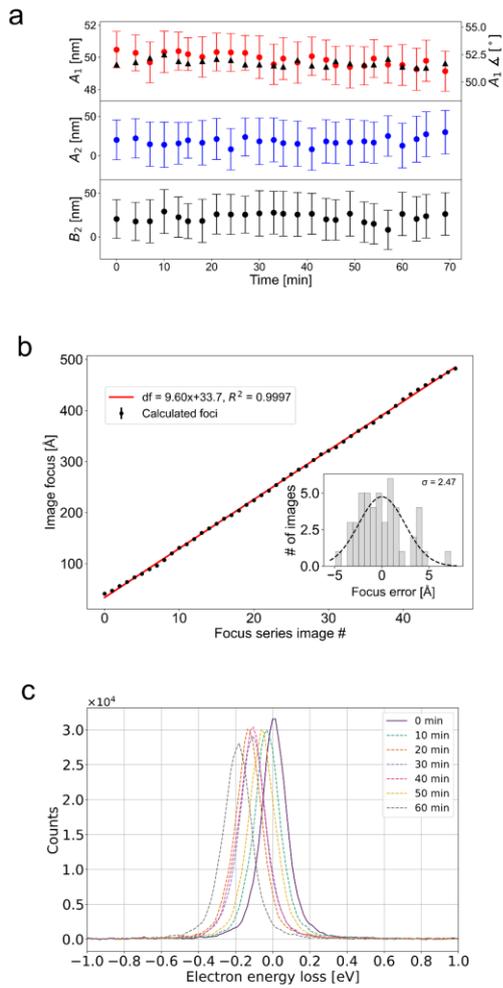

*Figure 1 – Time-resolved measurements of electron optical parameters. (a) Two-fold astigmatism ($A_1$), three-fold astigmatism ($A_2$), and coma ($B_2$) measured every three minutes over a 70 min period. $A_1$ angle is presented with a black triangle mark. The measurement confidence level is represented by the error bars. (b) Focus values for images in a focal series acquired with 47 images starting at nominal image focus df = 2.5 nm and with a nominal focus step of 1 nm. The best linear fit to the measured is indicated by the red line, where df is the image focus (in Å), x is the frame number, and $R^2$ is the coefficient of determination for the linear fit. The deviation between individual image focus values and the linear fit is defined as the focus error. The insert shows the distribution of focus errors. A Gaussian fit (superimposed black dashed line) has a mean value of 0 Å and a standard deviation of 2.47 Å. (c) Electron energy zero loss peak measured in 10-min intervals for 1 hr for a direct transmitted electron beam.*

Next, $N_2$ is introduced into the open gas-cell. **Figures 2a, b** show a low-loss electron energy loss (EEL) spectrum and a core-loss EEL spectrum at the N $L_{2,3}$ ionization edge, matching an earlier report.[26] Moreover, the HRTEM image intensity drops on the direct electron detector by ca. 6% at 1 mbar of $N_2$ as compared to 0 mbar $N_2$, consistent with a previous examination of a similar open gas setup (**Fig. 2c**).[6]

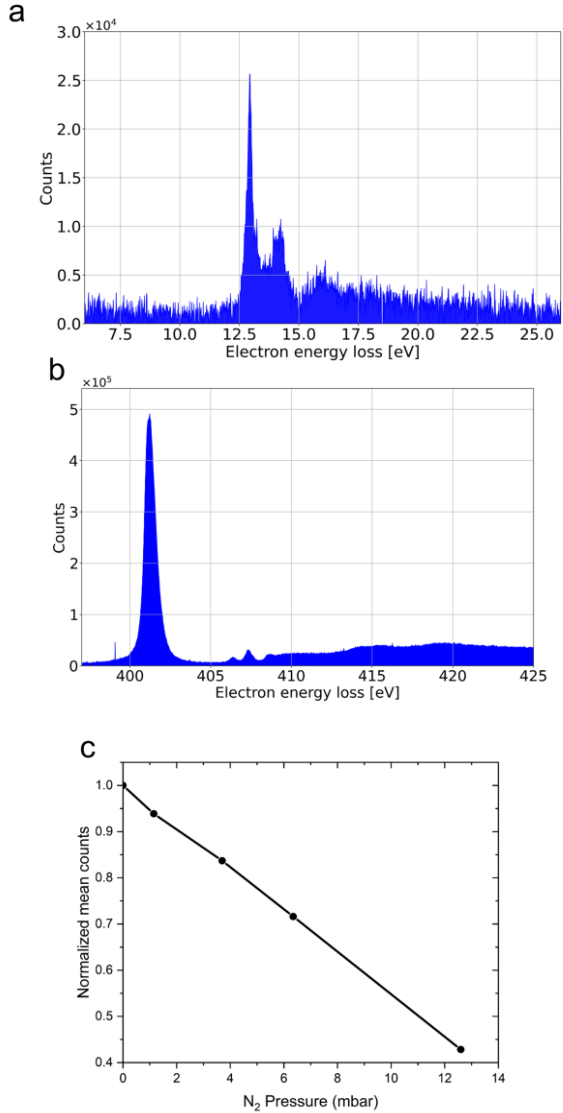

*Figure 2 – Characteristics of $N_2$ in the open gas-cell. (a) Electron energy low-loss spectrum of $N_2$ and (b) electron energy core-loss spectrum at the N $L_{2,3}$ ionization edge at ~ 401 eV. Both (a) and (b) are acquired after zero-loss-peak alignment to 0 eV. No background subtraction and any other data processing was done on the presented data. (c) Mean count of HRTEM images acquired using the direct electron detector as a function of increasing $N_2$ pressure with a fixed incident dose-rate of 1900 e/Å²/sec and an image acquisition time of 1 sec. The counts are normalized to the mean counts on the detector without $N_2$. while the beam traversed only the gas phase region.*

The electron optical stability of VISION PRIME heralds a high spatial resolution. Young fringe experiments were performed with dose-rates of ca. 1100-1200 e/Å²/sec to suppress the electron-beam interaction with the sample and the gaseous environment.[6,7] Moreover, all images are acquired in Nelsonian illumination mode[10] to restrict the physical extension of the electron illumination to minimize electron beam-gas interactions known to impair the resolution.[7] **Figures 3a, b** show the YF image and its corresponding FFT at 0 mbar $N_2$. **Figure 3c** is similar to F**ig. 3b** but with an orthogonal image shift to demonstrate isotropy in the YF measurements. That is, **Figs. 3b, c** reveal the information limit of 50 pm, as determined by the interference fringes in the FFTs. **Figures 3d-f** show the YF experiments at 1 mbar of $N_2$ (similar to **Figs. 3a-c**). Importantly, **Figs. 3e, f** also reveal a 50 pm information limit for 1 mbar $N_2$ present in the open gas-cell. These findings aligns with previous studies in the instrument-limited resolution limit of ~ 1Å [6,7]. Herein, 1 mbar $N_2$ was considered as an additional amorphous film of a few nm thickness only that causes a minor decrease in the transmittance, and, that excites beam-gas-sample interactions resulting in structural blurring and deterioration of resolution, unless low electron dose-rate illumination is applied.[6,7] Thus, this description seems to extend into the object-limited resolution regime in which the atomic potentials set the resolution.

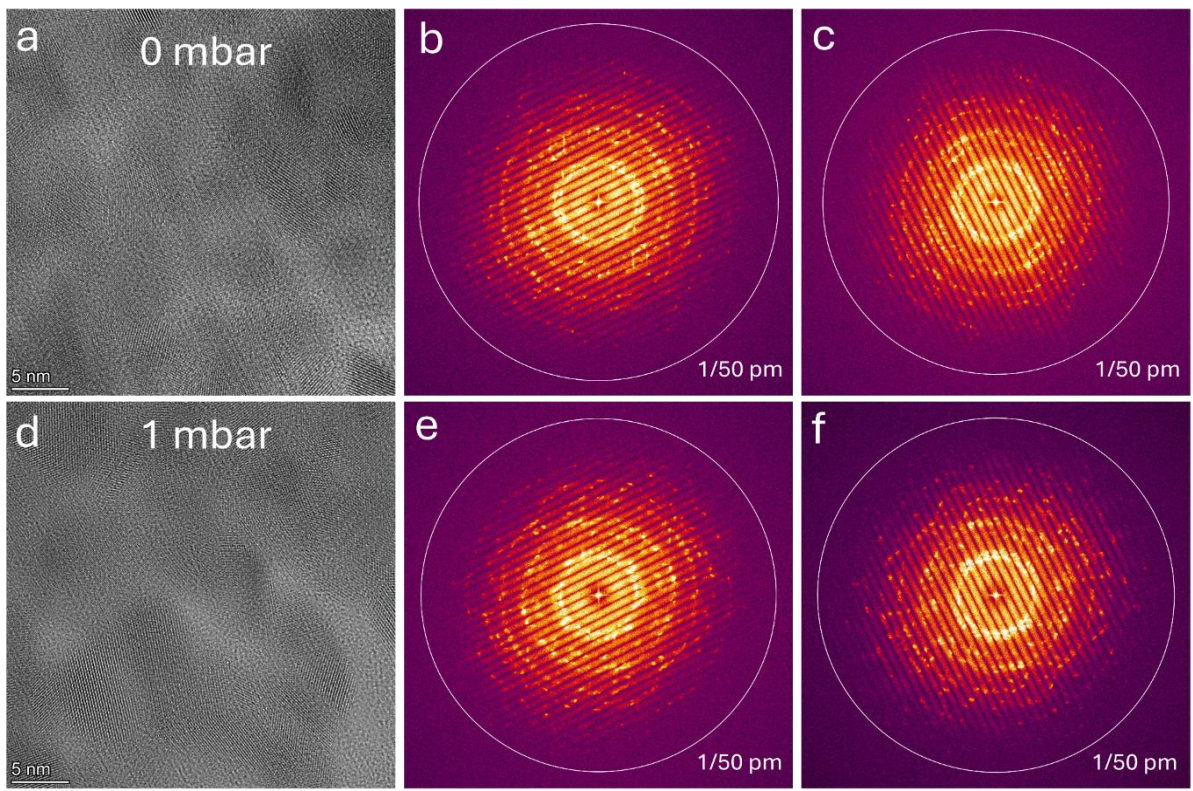

*Figure 3 – Information limit measurements at (a-c) 0 mbar and (d-f) 1 mbar of $N_2$. (a) Young fringe image, shifted in the x direction, acquired with a dose-rate of 1200 e/Å$^2$/s and a dose of 19200 e/Å$^2$. (b) Fast Fourier transform of (a). (c) Fast Fourier transform of YF image acquired as in (b) but shifted in the y direction, at 1100 e/Å$^2$/sec and dose of 18700 e/Å. (d) Young fringe image, shifted in the x direction, acquired with a dose-rate of 1100 e/Å$^2$/sec and a dose of 18700 e/Å$^2$.(e) Fast Fourier transform of (d). (f) Fast Fourier transform of YF image acquired as in (e) but shifted in the y direction, at dose-rate 1100 e/Å$^2$/sec and dose 18700 e/Å$^2$. In all FFT's of YF images, a white circle is superimposed with a radius indicated by the largest extend of the YF modulated contrast. The corresponding spatial frequency is indicated corresponding to the inverse information limit.*

Young fringe images are an accepted approach for evaluating the information limit, i.e. optical stability of an electron microscope, representing a reciprocal space measure of the entire field-of-view over the sample. Thus, it is considered a spatially invariant microscope resolution measurement. While this is certainly applicable for an instrument-limited resolution, it becomes questionable in the case of object-limited resolution. That is, nanostructured materials are associated with abundant surfaces and interfaces where atoms can differ in coordination and atomic arrangements from bulk sites. As a consequence, atoms' dynamic behaviours, including atom vibrations, can vary among surface, interface, and bulk sites, affecting the ultimate local spatial resolution.[4]

To address resolution in a localized and quantitative manner, the reconstructed complex exit wave images obtained at 0 mbar (**Figs. 4a-d**) and at 1 mbar $N_2$ (**Figs. 4e-h**) as analyzed with respect to the width of the column peaks in the phase image, as it measures the extend of the projected electrostatic potential of the atomic columns in a zone axis orientation.[27] For the reconstructions, the focal spread was set by the experimentally obtained information limit of $\rho_{info} = \left(\frac{\delta\lambda\pi}{2}\right)^{\frac{1}{2}} = 50 \ pm$ [28], which is corresponding to a value of

$\delta = \frac{2\rho_{info}^2}{\pi\lambda} = 8.1$ Å, which is within the focal spread theoretical upper limit. By comparing the width of the phase peak of neighboring atomic columns, insights into the locally varying object-limited resolution can be obtained.[4]

**Figures 4a, b** show the exit wave phase and modulus image of a carbon-supported polycrystalline Au sample at 0 mbar $N_2$. A single-crystal domain, ca. 4 nm wide, containing 320 atomic columns is outlined in the phase image by a dashed line. A FFT of the reconstructed exit wave in **Fig. 4c** shows spatial frequencies transferred up to 53 pm, potentially limited by the low signal-to-noise ratio (SNR) over a small field of view, as the particle in the image is cropped from the center of the full field-of-view. Specifically, the FFT of the exit wave in **Fig. 4c** reveals a pattern of reciprocal lattice vectors consistent with the orientation close to Au [110]. Moreover, the angle between Au [111] and Au [200] is ~ 57.8° degrees which is ~ 3° degrees off the [110] zone axis for Au in face-centered cubic (FCC) structure. The width of the atomic columns is determined by fitting a 2D Gaussian function to their phase profiles. **Fig. 4d** shows a close-up on a single phase peak (marked by a blue point in **Fig. 4a**) for which the Gaussian fit yields an ellipticity of ε=0.91, and a calculated equivalent diameter of 2<σ> = 49.6 pm. Similarly, **Fig. 4h** shows a close-up on a more elliptically shaped phase peak that was fitted with ε=0.85, and a calculated equivalent diameter 2<σ> = 60.2 pm (marked in red in **Fig. 4a**).

**Figures. 4e, f** show an exit wave phase and modulus image of a Au polycrystalline domain with a total of 293 atomic columns outlined by a dashed line, observed under exposure to 1

mbar $N_2$, consisting of three distinct single crystalline domains; each being ca. 2 nm - 2.5 nm wide in size. **Figure 4g** shows the corresponding FFT, revealing reciprocal lattice vectors consistent with the domains being oriented close to Au [110] and multiple lattice spots consisting of different in-plane domain rotations. The FFT in **Fig. 4g** indicates spatial frequencies corresponding to a real space resolution of 57 pm. However, since the image was cropped from the center of the full field-of-view the SNR is limited in the image and might inhibit transfer of higher-frequency spots in the FFT (as in **Fig. 4c**). In addition, based on reciprocal lattice vectors in the largest domain, the angle between Au [111] and Au [200] is ~ 56.5° degrees which is ~ 2° degrees off the [110] Au zone axis in the FCC structure.

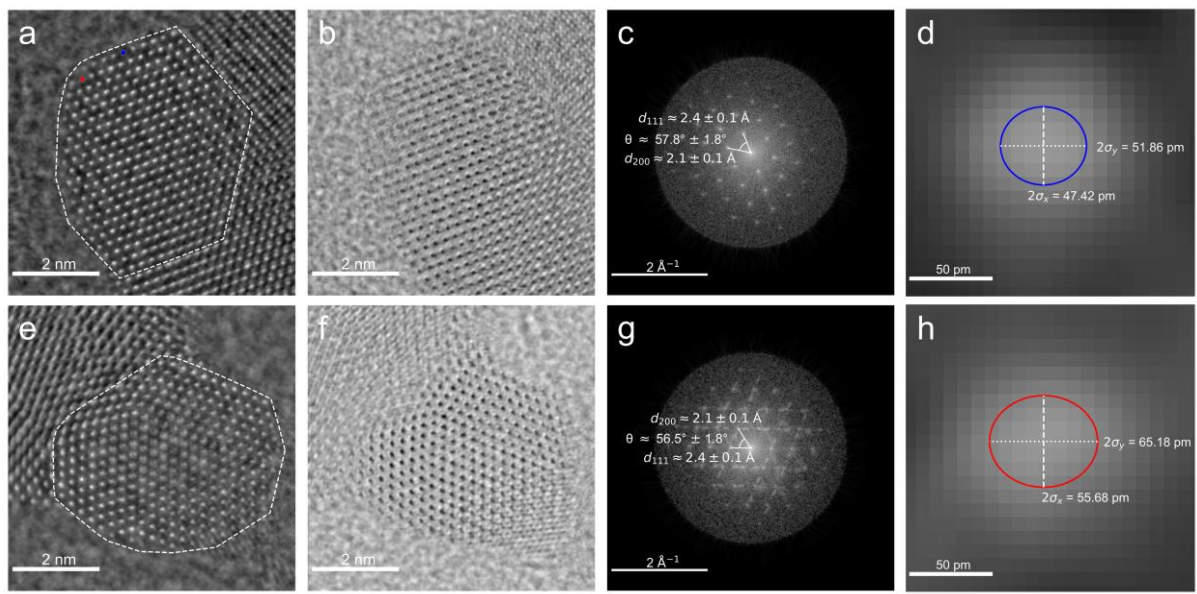

*Figure 4. Analysis of exit waves of Au crystalline domains. (a) Phase and (b) modulus image of an exit wave at 0 mbar of $N_2$. The reconstruction was prepared using the first 47 images of a focus series of images to match the 0 mbar YF dose conditions (focus calculation are seen in Fig. 1b). (c) Fast Fourier transform of the reconstructed exit wave, showing a zone axis close to Au [110]. The errors are estimated by a pixel change in measuring the distance and angle between Au [111] and Au [200]. (d) Close-up on a phase peak marked blue in panel (a), including an atomic column with an ellipticity value of $\varepsilon$=0.91. (e) Phase and (f) modulus image of an exit wave at 1 mbar of $N_2$. (g) Fast Fourier transform of the reconstructed exit wave, showing a zone axis close to Au [110]. The errors are estimated as in panel (c). (h) Close-up on a phase peak marked red in panel (a) with $\varepsilon$=0.85. For panels (c) and (g), the FFTs are limited by the 1/(50 pm) aperture included in the exit wave reconstruction. All acquisitions were taken using the same pixel size and illumination conditions.*

**Figures 5 a-d** highlight the atomic columns in the exit wave phase image at 0 mbar (**Fig. 4a**) that are fitted with a 2D elliptical Gaussian function. Specifically, **Fig. 5b** show the ellipticity value $\varepsilon$ and **Fig. 5c** the atomic column width ($2<\sigma>$) of all the atomic columns of the marked crystalline Au domain. Noticeably, the ellipticity decreases from $\varepsilon$ = 1.0-0.8 near the surface terminations to $\varepsilon$ = 0.8-0.6 closer to the center of the nanoparticle, indicating the atomic columns in bulk of the crystalline domain have a more elongated shape compared to surface terminations. This effect may be attributed to a slight tilt, thickness effects, or defects in the Au particle.[29] Next, **Figs. 5e-h** show atomic columns of the Au sample in the exit wave phase image at 1 mbar $N_2$ (**Fig. 4e**) fitted with the 2D elliptical Gaussian function. **Figure 5f** indicates a potentially off-axis tilt or bending of the atomic columns at the grain boundaries, as the atomic columns in these positions appear wider and elongated compared to the rest of atomic columns.[30] In addition, these broadening effects are convoluted further by exit wave defocus effects. Defocus effects are introduced during local aberration correction and exit wave propagation, both optimized to the edge atomic columns of the Au particle. The column elongations in the center of the polycrystalline Au region are not further analyzed.

Next, we focus on the (111) surface terminations of the crystalline Au domains at 0 and 1 mbar $N_2$, respectively (**Figs. 5d, h**). Interestingly, in the high vacuum case there is no apparent difference in the atomic column widths between the columns at the surface termination (**Fig. 5d**). The columns' width averaged over the first atomic row is 62.5 pm, while the second row has an average of 62.4 pm. In contrast, under 1 mbar $N_2$ conditions, the atomic column width differs markedly between the first and second atomic row terminating the surface, with an average column width of the first atomic row being 68.1 pm and of the second row being 57.3 pm (**Fig. 5h**). The variations in column widths at the surface of a nanocrystalline domain could be affected by a limited precision in the fitting or contrast blurring due to thickness variations in the supporting carbon film. Nevertheless, the clear widening of the atomic columns by 10.8 pm (larger than the imaging pixel size of 8.3 pm/pixel) in the presence of $N_2$ could also indicate the onset of the beam-gas-sample interactions, previously reported to limit the instrument resolution to 100 pm in earlier implementations of open gas-cells with transmission electron microscopes.[7,31]

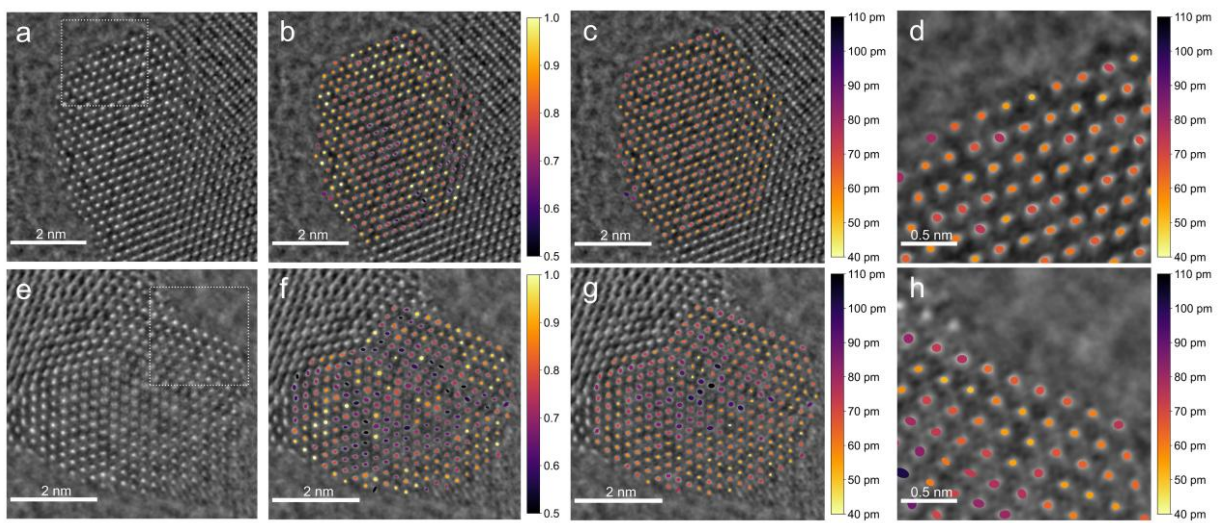

*Figure 5 – Analysis of atomic columns in exit wave phase images of nanocrystalline Au. (a,e) Phase image of the sample at 0 mbar and 1 mbar of $N_2$, respectively (same as Fig. 3a, e). (b,f) Ellipticity of the atomic columns at 0 mbar and 1 mbar, respectively, superimposed as the colored discs on the phase images in panels (a) and (e). (c,g) The $2<\sigma>$ atomic column diameters at 0 mbar and 1 mbar, respectively, superimposed as the colored discs on the phase images in panels (a) and (e). (d) Close-up on the (111) surface termination indicated by a white frame in panel (a) with the $2<\sigma>$ atomic column diameters superimposed with the colored discs. (h) Close-up on the (111) surface termination indicated by a white frame in panel (e) with the $2<\sigma>$ atomic column diameter superimposed with the colored discs.*

# Conclusions:

The present study shows that electron optics and open gas-cells can be combined to provide 50 pm resolution images of nanostructures under exposure to a few mbar gas pressures. In the present work, this resolution was obtained using Nelsonian illumination to match the field-of-view and field-of-illumination at low dose-rates, an image corrector optics capable of optical transfer down to 50 pm and a direct detection camera in counting mode ensuring the transfer of high frequencies.

Moreover, the stability of the electron-optical state of the new VISION PRIME system opens quantitative imaging modes such as electron exit wave reconstruction and its analysis to address the dynamic behavior of surface atoms at the single-atom level relevant for heterogeneous catalysis, crystal growth, and corrosion.

# Competing interests:

# Acknowledgments:

The authors acknowledge financial support for the Center for Visualizing Catalytic Processes (VISION) by the Danish National Research Foundation (DNRF146) and the Technical University of Denmark (DTU). We thank David Foord, Narashima Shastri, Dennis Cats, and Wessel Haasnoot from Thermo Fisher Scientific for an extraordinarily fruitful collaboration and valuable contributions to the design and establishment the VISION PRIME electron microscope.

# References:

[1] C. Kisielowski, B. Freitag, M. Bischoff, H. Van Lin, S. Lazar, G. Knippels, P. Tiemeijer, M. Van Der Stam, S. Von Harrach, M. Stekelenburg, M. Haider, S. Uhlemann, H. Müller, P. Hartel, B. Kabius, D. Miller, I. Petrov, E.A. Olson, T. Donchev, E.A. Kenik, A.R. Lupini, J. Bentley, S.J. Pennycook, I.M. Anderson, A.M. Minor, A.K. Schmid, T. Duden, V. Radmilovic, Q.M. Ramasse, M. Watanabe, R. Erni, E.A. Stach, P. Denes, U. Dahmen, Detection of Single Atoms and Buried Defects in Three Dimensions by Aberration-Corrected Electron Microscope with 0.5-Å Information Limit, Microsc. Microanal. 14 (2008) 469–477. https://doi.org/10.1017/S1431927608080902.

[2] R. Erni, M.D. Rossell, C. Kisielowski, U. Dahmen, Atomic-Resolution Imaging with a Sub-50-pm Electron Probe, Phys. Rev. Lett. 102 (2009) 096101. https://doi.org/10.1103/PhysRevLett.102.096101.

[3] F.-R. Chen, D. Van Dyck, C. Kisielowski, In-line three-dimensional holography of nanocrystalline objects at atomic resolution, Nat. Commun. 7 (2016) 10603. https://doi.org/10.1038/ncomms10603.

[4] F.-R. Chen, D. Van Dyck, C. Kisielowski, L.P. Hansen, B. Barton, S. Helveg, Probing atom dynamics of excited Co-Mo-S nanocrystals in 3D, Nat. Commun. 12 (2021) 5007. https://doi.org/10.1038/s41467-021-24857-4.

[5] J.R. Jinschek, S. Helveg, L.F. Allard, J.A. Dionne, Y. Zhu, P.A. Crozier, Quantitative gas-phase transmission electron microscopy: Where are we now and what comes next?, MRS Bull. 49 (2024) 174–183. https://doi.org/10.1557/s43577-023-00648-8.

[6] J.R. Jinschek, S. Helveg, Image resolution and sensitivity in an environmental transmission electron microscope, Micron 43 (2012) 1156–1168. https://doi.org/10.1016/j.micron.2012.01.006.

[7] M. Ek, S.P.F. Jespersen, C.D. Damsgaard, S. Helveg, On the role of the gas environment, electron-dose-rate, and sample on the image resolution in transmission electron microscopy, Adv. Struct. Chem. Imaging 2 (2016) 1–8. https://doi.org/10.1186/s40679-016-0018-x.

[8] J.F. Creemer, S. Helveg, G.H. Hoveling, S. Ullmann, A.M. Molenbroek, P.M. Sarro, H.W. Zandbergen, Atomic-scale electron microscopy at ambient pressure, Ultramicroscopy 108 (2008) 993–998. https://doi.org/10.1016/j.ultramic.2008.04.014.

[9] L.F. Allard, S.H. Overbury, W.C. Bigelow, M.B. Katz, D.P. Nackashi, J. Damiano, Novel MEMS-Based Gas-Cell/Heating Specimen Holder Provides Advanced Imaging Capabilities for *In Situ* Reaction Studies, Microsc. Microanal. 18 (2012) 656–666. https://doi.org/10.1017/S1431927612001249.

[10] P.C. Tiemeijer, M. Bischoff, B. Freitag, C. Kisielowski, Using a monochromator to improve the resolution in TEM to below 0.5 Å. Part I: Creating highly coherent monochromated illumination, Ultramicroscopy 114 (2012) 72–81. https://doi.org/10.1016/j.ultramic.2012.01.008.

[11] F. Zemlin, K. Weiss, P. Schiske, W. Kunath, K.-H. Herrmann, Coma-free alignment of high resolution electron microscopes with the aid of optical diffractograms, Ultramicroscopy 3 (1978) 49–60. https://doi.org/10.1016/S0304-3991(78)80006-0.

[12] J. Barthel, A. Thust, Quantification of the Information Limit of Transmission Electron Microscopes, Phys. Rev. Lett. 101 (2008) 200801. https://doi.org/10.1103/PhysRevLett.101.200801.

[13] M. Lentzen, Contrast Transfer and Resolution Limits for Sub-Angstrom High-Resolution Transmission Electron Microscopy, Microsc. Microanal. 14 (2008) 16–26. https://doi.org/10.1017/S1431927608080045.

[14] R.T.K. Baker, P.S. Harris, Controlled atmosphere electron microscopy, J. Phys. [E] 5 (1972) 793–797. https://doi.org/10.1088/0022-3735/5/8/024.

[15] E.D. Boyes, P.L. Gai, Environmental high resolution electron microscopy and applications to chemical science, Ultramicroscopy 67 (1997) 219–232. https://doi.org/10.1016/S0304-3991(96)00099-X.

[16] D.C. Bell, C.J. Russo, G. Benner, Sub-Ångstrom Low-Voltage Performance of a Monochromated, Aberration-Corrected Transmission Electron Microscope, Microsc. Microanal. 16 (2010) 386–392. https://doi.org/10.1017/S1431927610093670.

[17] http://www.totalresolution.com.

[18] R.W. Gerchberg, W.O. Saxton, A practical algorithm for the determination of plane from image and diffraction pictures, Optik 35 (1972) 237–246.

[19] D. Van Dyck, M. Op De Beeck, A simple intuitive theory for electron diffraction, Ultramicroscopy 64 (1996) 99–107. https://doi.org/10.1016/0304-3991(96)00008-3.

[20] C. Kisielowski, P. Specht, D. Alloyeau, R. Erni, Q. Ramasse, E.M. Secula, D.G. Seiler, R.P. Khosla, D. Herr, C. Michael Garner, R. McDonald, A.C. Diebold, Aberration-corrected Electron Microscopy Imaging for Nanoelectronics Applications, in: AIP Conf. Proc., AIP, Albany (New York), 2009: pp. 231–240. https://doi.org/10.1063/1.3251226.

[21] C. Kisielowski, R. Erni, B. Freitag, Object-defined Resolution Below 0.5Å in Transmission Electron Microscopy - Recent Advances on the TEAM 0.5 Instrument, Microsc. Microanal. 14 (2008) 78–79. https://doi.org/10.1017/S143192760808759X.

[22] C. Kisielowski, C.J.D. Hetherington, Y.C. Wang, R. Kilaas, M.A. O'Keefe, A. Thust, Imaging columns of the light elements carbon, nitrogen and oxygen with sub Ångstrom resolution, Ultramicroscopy 89 (2001) 243–263. https://doi.org/10.1016/S0304-3991(01)00090-0.

[23] M. Nord, P.E. Vullum, I. MacLaren, T. Tybell, R. Holmestad, Atomap: a new software tool for the automated analysis of atomic resolution images using two-dimensional Gaussian fitting, Adv. Struct. Chem. Imaging 3 (2017) 9. https://doi.org/10.1186/s40679-017-0042-5.

[24] C. Hetherington, Aberration correction for TEM, Mater. Today 7 (2004) 50–55. https://doi.org/10.1016/S1369-7021(04)00571-1.

[25] J. Barthel, A. Thust, On the optical stability of high-resolution transmission electron microscopes, Ultramicroscopy 134 (2013) 6–17. https://doi.org/10.1016/j.ultramic.2013.05.001.

[26] P.A. Crozier, S. Chenna, In situ analysis of gas composition by electron energy-loss spectroscopy for environmental transmission electron microscopy, Ultramicroscopy 111 (2011) 177–185. https://doi.org/10.1016/j.ultramic.2010.11.005.

[27] S. Van Aert, P. Geuens, D. Van Dyck, C. Kisielowski, J.R. Jinschek, Electron channelling based crystallography, Ultramicroscopy 107 (2007) 551–558. https://doi.org/10.1016/j.ultramic.2006.04.031.

[28] A.F. De Jong, D. Van Dyck, Ultimate resolution and information in electron microscopy II. The information limit of transmission electron microscopes, Ultramicroscopy 49 (1993) 66–80. https://doi.org/10.1016/0304-3991(93)90213-H.

[29] D.J. Smith, L.A. Bursill, G.J. Wood, Non-anomalous high-resolution imaging of crystalline materials, Ultramicroscopy 16 (1985) 19–31. https://doi.org/10.1016/S0304-3991(85)80004-8.

[30] T.P. Bartel, C. Kisielowski, A quantitative procedure to probe for compositional inhomogeneities in In$_x$Ga$_{1−x}$N alloys, Ultramicroscopy 108 (2008) 1420–1426. https://doi.org/10.1016/j.ultramic.2008.04.096.

[31] S. Helveg, C.F. Kisielowski, J.R. Jinschek, P. Specht, G. Yuan, H. Frei, Observing gas-catalyst dynamics at atomic resolution and single-atom sensitivity, Micron 68 (2015) 176–185. https://doi.org/10.1016/j.micron.2014.07.009.